\documentclass[3p,times]{elsarticle}

\usepackage{ecrc}

\volume{00}

\firstpage{1}

\journalname{Nuclear Physics A}

\runauth{J. Yamagata-Sekihara et al.}

\jid{NUPHA}

\jnltitlelogo{Nuclear Physics A}

\CopyrightLine{2012}{Published by Elsevier Ltd.}

\usepackage{amssymb}

\usepackage[figuresright]{rotating}

\begin{document}

\begin{frontmatter}

\dochead{}



\title{Nuclear density probed by Kaon-Nucleus systems and Kaon-Nucleus interaction}

\author[a]{J. Yamagata-Sekihara}\author[b]{N. Ikeno}\author[b]{H. Nagahiro}\author[c]{D. Jido}\author[b]{S. Hirenzaki}

\address[a]{Institute of Particle and Nuclear Studies, High Energy Accelerator Research Organization (KEK), Ibaraki 305-0801, Japan}
\address[b]{Department of Physics, Nara Women's University, Nara 630-8506, Japan}
\address[c]{Yukawa Institute for Theoretical Physics, Kyoto University, Kyoto 606-8502, Japan}

\begin{abstract}
Effective nuclear densities probed by kaon- and anti-kaon-nucleus systems are studied theoretically both for bound and low energy scattering states.
As for the anti-kaon bound states, we investigate kaonic atoms. We find that the effective density depends on the atomic states significantly
and we have the possibility to obtain the anti-kaon properties at various nuclear densities by observing the several kaonic atom states. 
We also find the energy dependence of the probed density by kaon and anti-kaon scattering states.
We find that the study of the effective nuclear density will help to find the proper systems to investigate the meson properties at various nuclear densities.
\end{abstract}


\end{frontmatter}


\section{Introduction}
\label{intro}
Meson-Nuclear systems are very important and useful objects to extract the meson properties at finite density, 
which may have close connections to the chiral symmetry breaking and its partial restoration in the nucleus \cite{Suzuki:2002ae}.
The effective nuclear densities probed by pionic atoms are known to be around $\sim 0.6\rho_0$ for almost every pionic state~\cite{Seki:1983sh,Ikeno:2011mv}
and the change of the chiral order parameter $\langle{\bar q}q\rangle$ in nuclei is only determined at this specific density~\cite{Suzuki:2002ae}.
Thus, we are very interested in the density probed by various meson-nucleus systems to find out the proper mesonic systems to explore the meson properties at various nuclear densities, which will be important to investigate the aspects of QCD symmetries at various nuclear densities beyond the linear density approximation~\cite{Jido:2011db}.

\section{Formulation}
\label{form}
We calculate the effective density of bound states for anti-kaon (kaonic atoms), and scattering states of anti-kaon and kaon with nucleus.
As for the kaonic atoms, we already do the similar analysis as reported in Ref.~\cite{Yamagata:2006sv}, where 
we define the overlapping probability ($S_{\rm bound}(r)$) of the kaonic atom density ($|R(r)|^2$) with the nucleon density distribution ($\rho(r)$) to obtain the effective density, as 
\begin{eqnarray}
S_{\rm bound}(r)&=&|R(r)|^2\rho(r)r^2~~.
\label{S_bound}
\end{eqnarray}
Since the integration of the $S_{\rm bound}(r)$ function gives the shift of anti-kaon binding energy as $\displaystyle \Delta{\rm B.E.}\propto\int d^3 r S_{\rm bound}(r)$ in the first order perturbation theory for the change of the potential term depending on the density linearly,
we define the nucleon density at the peak position of $S_{\rm bound}(r)$ as the effective nuclear density for bound states, which is expected to be most sensitive part of $\rho$ to the meson binding energy.

For scattering states, we newly determine the different overlapping probability ($S_{\rm scat}(r)$) as,
\begin{eqnarray}
S_{\rm scat}(r)&=&\sum_l (2l+1)|R_l(r)|^2\rho(r)r^2P_l(\cos(\theta_0))~~.
\end{eqnarray}
In this case, we define the $S_{\rm scat}(r)$ function so as to have the peak at the density which is most sensitive to a differential cross section, namely the change of the differential cross section is related to the $S_{\rm scat}(r)$ as, $\displaystyle \Delta\Bigl(\frac{d\sigma}{d\Omega}\Bigr)\propto\int d^3 r S_{\rm scat}(r)$ instead of the binding energy for the bound states. 
Here, we consider the scattering with $^{12}$C at $\theta_0=0$.

\begin{figure}[htb]
\begin{center}
\includegraphics[width=.80\textwidth]{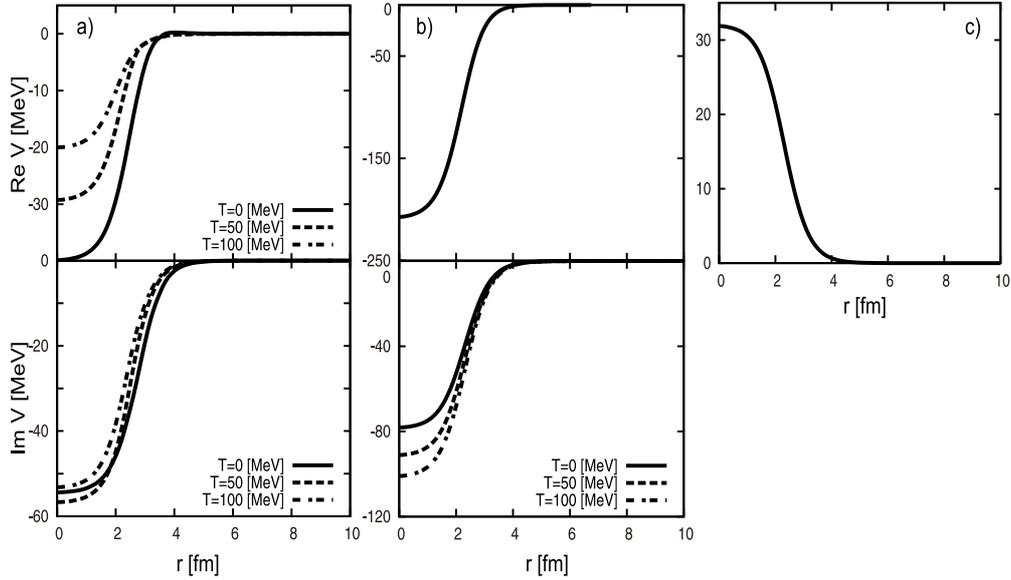}
\caption{Anti-kaon optical potentials for $^{12}$C obtained by (a) chiral unitary approach~\cite{Ramos:1999ku} and 
(b) phenomenological fit \cite{Batty:1997zp,Mares:2004pf} are shown as a function of the radial coordinate $r$.
The kaon-$^{12}$C potential (c) obtained in Ref.~\cite{Oset:2000eg} is also shown.
The upper and lower panels show the real and imaginary part of the potential, respectively. The kaon-nucleus potential in Ref.~\cite{Oset:2000eg} doesn't have imaginary part.
Some of the potential have the energy dependence as shown in the figures.}
\label{fig1}
\end{center}
\end{figure}

The meson wavefunction $R(r)$ in Eqs.~(1) and (2) are obatined by solving the Klein-Gordon equation with meson-nucleus optical potential.
In this article, we use two kinds of the optical potentials for anti-kaon, which are the chiral unitary potential~\cite{Ramos:1999ku} and the phenomenological potential
~\cite{Batty:1997zp,Mares:2004pf}.
For $K^+$ scattering, we use the potential given in Ref.~\cite{Oset:2000eg}.
The all potentials used in this article are shown in Fig.~\ref{fig1}.

\section{Numerical Results}
\label{results}
\begin{figure}[htb]
\begin{center}
\includegraphics[width=.5\textwidth]{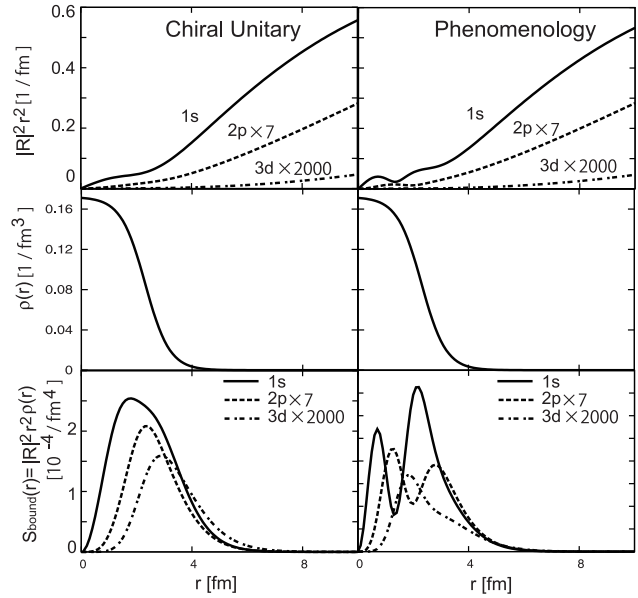}
\caption{Overlapping probabilities (lower frame) of the anti-kaon wave functions (upper frame ) with the nucleon densities (middle frame) in the atomic kaon bound states in $^{12}$C calculated with the (left)
 chiral unitary and (right) phenomenological optical potentials.}
\label{fig2}
\end{center}
\end{figure}

As for the kaonic atoms, we show the calculated anti-kaon densities ($|R(r)|^2$) and the nucelon density distributions ($\rho(r)$), and the overlapping probabilities ($S_{\rm bound}(r)$) in Fig.~\ref{fig2} \cite{Yamagata:2006sv}.
The nuclear density at the peak position of $S_{\rm bound}(r)$ shown in the lower panels corresponds to the effective nuclear density.
We can see that the effective densities probed by atomic kaons depend on the atomic orbits significantly.
Comparing the results of two potentials, we find the differences of the structure of the overlapping probabilities $S_{\rm bound}(r)$.
These differences are related to the properties of the deeply bound kaonic nuclear state for both potential cases, since
the kaonic nuclear states exist for both potentials.
In the phenomenological potential, the kaonic nuclear states are deeper than those in the chiral unitary potential. 
The wave functions of the atomic and nuclear states should be orthogonal.
Owing to the orthogonality the wave functions of atomic states have nodes in the region of the nuclear size~\cite{Yamagata:2005ic}.
The effective densities obtained from the overlapping probabilities will have information on the difference of the potentials, which might be extracted from experimental data.

\begin{figure}[htb]
\begin{center}
\includegraphics[width=.80\textwidth]{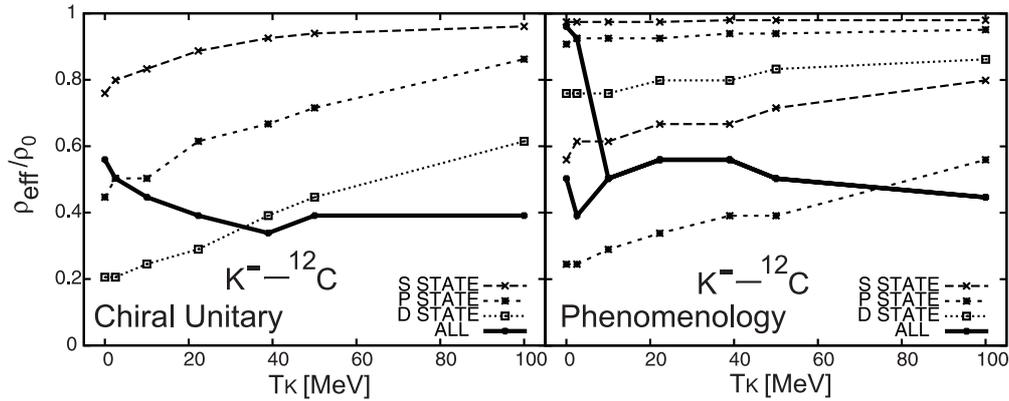}
\caption{Calculated effective nuclear densities defined in Eq.~(2) (solid line) of anti-kaon scattering for $^{12}$C with chiral unitary approach (left) and 
phenomenological fit (right) as plotted as the functions of the anti-kaon kinetic energy [MeV]. The contributions from a first few partial waves in Eq.~(2) are also shown in the figure.} 
\label{fig3}
\end{center}
\end{figure}

\begin{figure}[htb]
\begin{center}
\includegraphics[width=.45\textwidth]{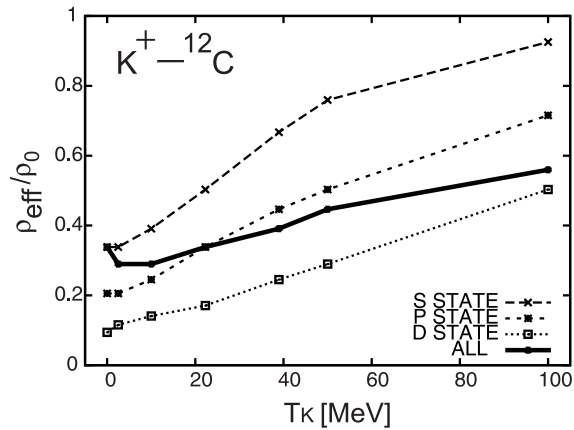}
\caption{Same as Fig.~\ref{fig3} except for the kaon scattering case.} 
\label{fig4}
\end{center}
\end{figure}

For scattering states, we show the calculated results of the energy dependence of the effective density in Figs.~\ref{fig3} (anti-kaon) and \ref{fig4} (kaon).
From these results, we find the effective density for anti-kaon dicreases for higher anti-kaon energy.
It would be natural that the effective density for higher energy could become smaller
than that for lower energy since the important partial waves become larger for higher energy
in general, and the wave function with larger angular momentum will observe more outer
part of the nucleus due to the centrifugal barrier. This is the case for the anti-kaon scattering.
For the anti-kaon, the increase of the effective density for each partial wave is moderate,
because the anti-kaon exists inside the nucleus with small probability due to the large absorption.
In contrast, the effective density for kaon increases for higher kaon energy. This is because the potential
of kaon in Ref.~\cite{Oset:2000eg} is weak repulsive and does not have imaginary part. It means that the kaon with
higher energy can penetrate into the nucleus without any absorption. This makes the effective
density of each partial wave increase rapidly. This is the reason that the kaon system has different
effective densities from those of the anti-kaon system.

\section{Summary}
We have calculated the effective nuclear density probed by anti-kaon and kaon.
The effective nuclear density is defined from the overlapping probability of the meson density and the nucleon density distribution.
As for the bound states, we consider the kaonic atoms.
We find the effective density probed by anti-kaon in atomic orbit depends on the quantum numbers of the atomic states.
For the scattering states, we have investigated the energy dependence of the effective density.
From this study, we find the effective density of anti-kaon decreases for higher anti-kaon energy due to the centrifugal barrier and the strong absorption inside the nucleus.
We also find the kaon for higher energy could probe higher nuclear density, which is the opposite tendency to the anti-kaon system.








\end{document}